\documentclass[a4paper,11pt]{article}
\usepackage{pos}

\usepackage{float}
\usepackage{amsmath}
\usepackage{amssymb}
\usepackage{graphicx}
\usepackage{mathtools,slashed}
\usepackage{bm}
\usepackage{comment}
\usepackage{xcolor}
\usepackage{subfigure}
\usepackage{array}
\usepackage{multirow}
\usepackage{url}
\usepackage{color}
\usepackage[T1]{fontenc}
\usepackage{graphicx}
\usepackage{esint}
\usepackage{hyperref}
\usepackage{atbegshi}
\usepackage{lipsum}

\newcommand{\M}[0]{{\mathcal{M}}}
\newcommand{\T}[0]{{\mathcal{T}}}
\newcommand{\K}[0]{{\mathcal{K}}}
\newcommand{\cH}[0]{{\mathcal{H}}}
\newcommand{\bH}[0]{{\mathbf{H}}}

\title{Accessing scattering amplitudes using quantum computers}

\author[a,b]{Ra\'ul A.~Brice\~no}
\author[b]{Marco A. Carrillo}
\author*[a]{Juan V. Guerrero}
\author[c]{Maxwell T. Hansen}
\author[d]{Alexandru M. Sturzu}

\affiliation[a]{Thomas Jefferson National Accelerator Facility, 12000 Jefferson Avenue, Newport News, Virginia 23606, USA}
\affiliation[b]{Department of Physics, Old Dominion University, Norfolk, Virginia 23529, USA}
\affiliation[c]{Higgs Centre for Theoretical Physics, School of Physics and Astronomy, The University of Edinburgh, Edinburgh EH9 3FD, UK}
\affiliation[d]{Department of Physics, New College of Florida, 5800 Bay Shore Rd, Sarasota, FL 34243, USA}

\emailAdd{juanvg@jlab.org}

\abstract{Future quantum computers may serve as a tool to access non-perturbative real-time correlation functions. In this talk, we discuss the prospects of using these to study Compton scattering for arbitrary kinematics. The restriction to a finite-volume spacetime, unavoidable in foreseeable quantum-computer simulations, must be taken into account in the formalism for extracting scattering observables. One approach is to work with a non-zero $i \epsilon$-prescription in the Fourier transform to definite momentum and then to estimate an ordered double limit, in which the spacetime volume is sent to infinity before $\epsilon$ is sent to $0$. For the amplitudes and parameters considered here, we find that significant volume effects arise, making the required limit very challenging. We present a practical solution to this challenge that may allow for future determinations of deeply virtual Compton scattering amplitudes, as well as many other reactions that are presently outside the scope of standard lattice QCD calculations.}

\FullConference{%
 The 38th International Symposium on Lattice Field Theory, LATTICE2021
  26th-30th July, 2021
  Zoom/Gather@Massachusetts Institute of Technology
}

\begin{document}
\maketitle

\section{Introduction}

Understanding how quarks and gluons are distributed within hadrons remains an overarching goal of modern-day nuclear physics. Among the physical processes used to asses the internal structure of such states is Compton scattering, which has been proposed as a tool to obtain  generalized parton distributions of hadrons~\cite{Ji:1996ek,Radyushkin:1996nd}. For this reason, the Compton scattering process is particularly relevant in the 12 GeV upgrade at Jefferson Lab~\cite{Dudek:2012vr} as well as the future electron-ion collider~\cite{Accardi:2012qut}.

Lattice quantum chromodynamics (lattice QCD) is the only known systematically improvable method for making non-perturbative predictions based in the fundamental theory of the strong nuclear force. In order to be numerically tractable, lattice QCD calculations must be defined in a finite Euclidean spacetime, which inherently limits the classes of observables that are accessible. For example the Compton amplitude, together with a wide class of other scattering and decay amplitudes, requires physical, Minkowski time evolution. It can therefore only be accessed from Euclidean correlators via analytic continuation or finite-volume methods. While the first method has received considerable attention recently (see e.g.~Refs.~\cite{Tripolt:2018xeo,Bailas:2020qmv,Bulava:2021fre}), the second is well established and has proven very useful for extracting hadronic scattering and decay amplitudes~\cite{Luscher:1986pf,Hansen:2019nir,Briceno:2017max,Rusetsky:2019gyk}.

Another promising numerical approach for evaluating real-time QCD quantities involves using quantum computing techniques (for a review of these ideas see~\cite{Georgescu:2013oza,Banuls:2019bmf}, and for recent applications see~\cite{Jordan:2014tma, Jordan:2011ci, Jordan:2011ne, Davoudi:2019bhy, Kuno:2014npa,Martinez:2016yna,Mueller:2019qqj,Lamm:2019uyc, Kaplan:2018vnj, Kaplan:2017ccd, Gustafson:2019mpk, Marshall:2015mna,Lu:2018pjk,Ciavarella:2020vqm}).
In this work, we discuss the prospects of accessing Compton-like amplitudes from Minkowski correlation functions. We define the finite-volume Minkowski correlator with a non-infinitesimal $i \epsilon$ (implemented via the Fourier transform) and note that the desired infinite-volume amplitude is given by the ordered double limit: $L \to \infty$ (where $L$ is the spatial periodicity) followed by $\epsilon \to 0$ (see also Refs.\cite{Hansen:2015zga,Bulava:2019kbi,Agadjanov:2016mao}). We then use the finite-volume formalism derived in~\cite{Briceno:2019opb} to predict the size of finite-volume systematic errors for given values of $\epsilon$ and $L$.

We additionally provide a prescription, based on averaging over redundant kinematics, that significantly reduces the finite-volume errors. For the theories that we consider this typically reduces the finite-volume effects by several orders of magnitude. The expectation is that the improvement provided by this procedure is universal. Although a proof remains outstanding, here we provide empirical evidence supporting this conjecture. The first evidence, published in Ref.~\cite{Briceno:2020rar}, showed that this procedure reduces finite-volume effects for kinematics where a single two-particle channel is kinematically open. In addition to reviewing these findings, in the present work we show preliminary results demonstrating that the same conclusion may be drawn for kinematics where multiple two-particle channels may are open.

\section{Infinite and finite volume amplitudes in 1+1D}

\begin{figure}[t!]
\begin{center}
\includegraphics[width=1\textwidth]{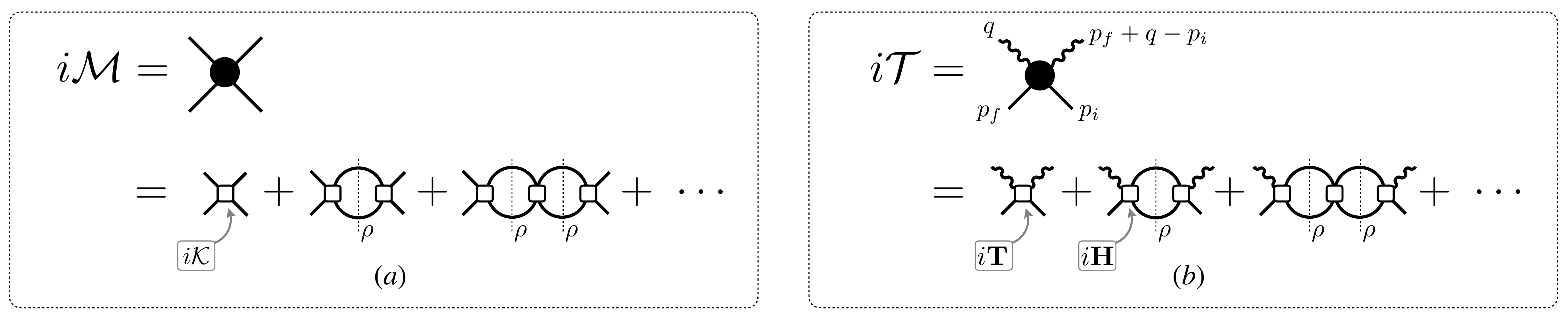}
\caption{Diagrammatic representations of $(a)$ the $2\to2$ scattering amplitude, $\M$, and $(b)$ the Compton-like amplitude, $\T$. \label{fig:iM_iT}}
\end{center}
\end{figure}

Ultimately, we are interested in the study of Compton scattering for arbitrary kinematics. However, at this stage, the formalism needed to describe the finite-volume artifacts for arbitrary kinematics has yet to be developed. Therefore, as a start, we restrict our attention to the kinematic region where two particle states may go on-shell. 

As shown in Ref.~\cite{Briceno:2019opb}, the kinematic singularities and finite-volume effects of these amplitudes are parametrized in terms of the infinite-volume amplitudes describing all physical subprocesses. Thus, in the present context, in order to understand the behavior of the Compton amplitude, we must first understand the $2\to2$ and $1+\mathcal{J}\to 2$ scattering amplitudes. The on-shell representation of these amplitudes are well known in 3+1D [see, for example,  Ref.~\cite{Briceno:2020vgp} for a recent detailed derivation]. Given that the first quantum computations are most likely to be performed in 1+1D, here we only consider 1+1D theories. In Ref.~\cite{Briceno:2020rar}, we first derived the expressions for these amplitudes, including the  Compton-like amplitude and its finite-volume analogue, in which a non-zero $\epsilon$ parameter is introduced in the definition. In what follows we quickly review the main necessary results.


\subsection{Amplitudes for a single two-particle channel in 1+1D}

We begin by considering the scattering of two hadrons of mass $m$ in 1+1D. The two-vector $P^\mu = (E, \boldsymbol P)$ denotes the total energy and momentum of the two-hadron state. In the center of mass frame the total energy ($E^\star$) is given by,
\begin{equation}
E^{\star2} = P^\mu P_\mu = E^2-\boldsymbol P^2 = s\,,
\end{equation}
where $s$ is the Mandelstam variable. In this section, we restrict our discussion to energies where two-particle systems may go on-shell. First, we assume there is a single two-particle channel open that is composed of two identical particles of mass $m$. This then implies that the energies considered will satisfy the condition $2m < E^\star < 3m$. We will later partly lift this assumption by allowing for multiple two-particle channels to dynamically couple. 

The $2\to2$ hadronic amplitude, denoted by $\M $, is defined diagrammatically in Figure~\ref{fig:iM_iT}(a). By isolating the singularities and summing these contributions to all orders, the amplitude can be written as 
\begin{align}
\M(s)
=\frac{1}{\K(s)^{-1}-i\rho(s)} \,,
\label{eq:Mdef}
\end{align}
where the the K-matrix, $\K $, is a real quantity for $s > 4 m^2$; and $\rho $ is the phase space factor. For identical particles in 1+1D the phase-factor is $\rho(s) = \frac{1 }{8 E^\star q^\star}$, where $q^\star$ is the relative momentum in the center-of-mass frame, $q^\star \equiv \sqrt{s/4 - m^2}$.

In Figure~\ref{fig:iM_iT}(a), it is possible to replace one of the initial hadron states with an external current. In this case, the 
$1+\mathcal{J}\to 2$ transition amplitude at all orders, analogous to $\M$, can be written as,
\begin{align}
\mathcal{H}(s)
=\M(s) \mathcal{A}(s ,Q^2)  \,,
\label{eq:calHdef}
\end{align}
where $\mathcal{A}(s ,Q^2)$ is a {\it generalized} transition form factor and is a smooth function in $s$~\cite{Briceno:2020vgp}.

Now, we consider the presence of two external currents to study Compton-like amplitudes. In particular, we focus on matrix elements involving the time ordered product, of two identical scalar currents, $\mathcal J(x)$, between an initial and final single-particle external state,
\begin{equation}
\T(s,Q^2,Q^2_{if}) \equiv i \int d^2 x \, e^{i\omega t - i\bm{q} \cdot \bm{x}} \, \langle \boldsymbol p_f \vert \, \text{T} \{ \mathcal J (x) \mathcal J'(0) \} \, \vert \boldsymbol p_i \rangle_{\text{c}}\,,
\label{eq:TAdef}
\end{equation}
where the subscript ``$\text{c}$'' indicates that the definition of $\mathcal{T}$ includes only connected contributions, and $\text{T}$ denotes the time ordering. The process kinematics is defined in the first line of Figure~\ref{fig:iM_iT}(b), where $q=(\omega, \bm{q})$. The Lorentz invariants relevant for this process are the Mandelstam variable $s = (p_f+q)^2$, 
while $Q^2_{if}=-(p_f+q-p_i)^2$ and $Q^2=-q^2$ are the incoming and outgoing current virtualities, respectively. 

The Compton-like amplitude $\mathcal{T}$ is diagrammatically defined in Figure~\ref{fig:iM_iT}(b). 
By isolating the possible singularities associated with intermediate two-particle states, one can write this amplitude in terms of $\M$, $\mathcal{A}$, and a new smooth real function, $\mathbf{S} $~\footnote{In Ref.~\cite{Briceno:2020rar} $\T$ was written explicitly in terms of $ \mathbf{T}$ and $\bH$. Here we instead use the equivalent expression obtained following the steps sketched in Ref.~\cite{Briceno:2020vgp,Briceno:tbp}},
\begin{align}
\mathcal{T} (s,Q^2 ,Q^2_{if})
&= \mathbf{S}(s,Q^2 ,Q^2_{if}) 
+ \mathcal{A}(s ,Q^2)\mathcal{M}(s)\mathcal{A}(s ,Q^2_{if}) +
[s\longleftrightarrow u]
\,
 \,,
\label{eq:compton}
\end{align}
where $[s\longleftrightarrow u]$ denotes the exchange of the Mandelstam variables $s$ by $u$.

Having established the relevant expressions for the Compton-like amplitude $\mathcal T $, we now focus on a finite-volume estimator for this quantity, $\mathcal T_L$, defined as 
\begin{equation}
\T_L(p_f, q, p_i) \equiv  2i \sqrt{\omega_{\boldsymbol p_f} \omega_{\boldsymbol p_i}} \, L\int dx^0 \int_0^L dx^1 \, e^{i \omega x^0 -\epsilon \vert x^0 \vert -i \bm{q} \cdot \bm{x}}\, \langle \boldsymbol p_f \vert \, \text{T} \{ \mathcal J(x) \mathcal J'(0) \} \, \vert\boldsymbol p_i \rangle_{\text{c}, L} \,,
\label{eq:TAdefFV}
\end{equation}
where the proportionality factor arises from the normalization of one-particle states in a finite volume. The $\epsilon$ regulates the singularities both in the $s$- and $u$-channel diagrams. Here we only consider explicitly this effect in the $s$-channel diagrams, where this shift can be understood as a shift in $q_0\to q_0+i\epsilon$. In going forward, we will simply ignore the contribution from the $u$-channel diagrams.

In order to understand how to recover the infinite volume amplitude, $\T$, from its finite-volume counterpart, $\T_L$, we consider the finite-volume long range formalism derived in Ref.~\cite{Briceno:2019opb} for 3+1D. The finite-volume analog for the Compton-Scattering amplitude, $\mathcal T_L$, can be written in terms of the infinite volume amplitudes and one finite-volume function, $F$. Ignoring exponentially suppressed volume effects, one finds  
\begin{align}
\T_{L}(p_f,q,p_i) =
\T (s,Q^2 ,Q^2_{if}) -
\cH(s,Q^2)
\,
\frac{1}{F^{-1}(E^\star,\boldsymbol P, L) + \M(s) } 
\,
\cH(s ,Q^2_{if})
\,,
\label{eq:TL}
\end{align}
and  one can show that in $1+1$D the $F$ function can be written as
\begin{align}
F(E^\star,\boldsymbol P, L )
&= i \rho(s)
+ 
\frac{\rho(s)}{2}
\left[
\cot\left(\frac{L\gamma(q^\star+\omega_q^\star\beta)}{2}\right)
+
\cot\left(\frac{L\gamma(q^\star-\omega_q^\star\beta)}{2}\right)
\right] 
\,,
\label{eq:Fcot}
\end{align}
where $\gamma$ and $\beta$ define a Lorentz boost in the $\boldsymbol P$ direction, $\gamma = E/E^\star$ and $\beta = \boldsymbol P /E$, and $\omega_q^\star=\sqrt{q^{\star 2}+m^2} = E^\star/2$.

It is easy to show that $F$ satisfies, 
\begin{align}
\lim_{\epsilon\to 0} \lim_{L \to \infty}F(E^\star +i\epsilon,\boldsymbol P, L )= 0 \, .
\label{eq:F_lim}
\end{align}
Thus the physical Compton-like amplitude can be recovered from the ordered double limit: 
\begin{align}
\lim_{\epsilon \to 0} \lim_{L \to \infty} \T_{L}(p_f,q,p_i) = \T (s,Q^2 ,Q^2_{if}) \,.
\label{eq:IV_lim}
\end{align}
In practice, one cannot take this limit numerically. Instead, one must resort to determining an estimate from various values of $\epsilon$ and $L$ and assigning a systematic uncertainty due to the non-zero and finite values, respectively, or due to the extrapolation ansatz.

\subsection{Extension for multiple open channels}

For energies where $n$ two-body channels may interact, the infinite volume scattering amplitude can be written as~\cite{Briceno:2012yi,Hansen:2012tf} 
\begin{align}
\mathcal{M}_{ab}(s)
    = \left[\left( 1 - i\mathcal{K}(s)\rho(s) \right)^{-1}\right]_{ab'}\mathcal{K}_{b'b}(s),
  \end{align}
where the indexes run over the possible channels. The K-matrix now is a square matrix of dimension $n\times n$, and  $\rho$ is a diagonal matrix defined by $\rho_{ab}(s) = \frac{\delta_{ab}}{8E^\star q_a^\star}$.  For simplicity, we consider that the two particles in each given channel are identical with mass $m_a$, and as a result the relative momentum for the $a$-th channel can be written as $q_a^\star \equiv \sqrt{s/4 - m_a^2}$. We choose $m_1$ to be the mass of the lightest particle, $m_1<m_{a\neq1}$. 

The infinite volume Compton amplitude remains a scalar quantity, 
\begin{align}
\mathcal{T}(s,Q^2,Q^2_{if}) 
    &= \mathbf{S}(s,Q^2,Q^2_{if}) 
       + \mathcal{A}_a(s,Q^2)\mathcal{M}_{ab}(s)\mathcal{A}_b(s,Q_{if}^2), 
\end{align}
where now the transition form factors, $\mathcal{A}$, are vectors in  channel space, and $\mathbf{S} $ still remains a scalar smooth function. For the finite volume Compton amplitude with multiple open channels, 
\begin{align}
\T_{L}(p_f,q,p_i)
    &= \mathcal{T}(s,Q^2,Q^2_{if}) 
       - \mathcal{H}_a(s,Q^2) \left[\left(F^{-1}(E^\star,\boldsymbol P, L )+\mathcal{M}(s)\right)^{-1}\right]_{ab}
         \mathcal{H}_b(s,Q^2_{if}),
\label{eq:TL_coupled}
\end{align}
the transition amplitudes, $\mathcal{H}$, are vectors in channel space  which may be expressed as $\mathcal{H}_a  = \mathcal{M}_{ab}  \mathcal{A}_b $ and $F$ is a finite-volume diagonal matrix whose elements are the geometric functions for each channel, as given in Eq.~\eqref{eq:Fcot} with the appropriate relative momentum  $q_a^\star$.

\section{Numerical Results}
\label{sec:Numerical_results}

In this section, we explore how to numerically recover the infinite-volume Compton amplitude, $\T$, from its finite-volume analog, $\T_L$. To achieve this, one requires to choose reasonable functional forms for the infinite-volume real functions $\K$, $\mathcal{A}$ and $\mathbf{S}$, since they enter in $\T_L$.
We will first discuss the results for a single channel open, and then consider multiple open channels.

\subsection{Single channel open}
\label{sec:single_num}

We use a flexible parametrization of the K matrix,
\begin{gather}
\K (s) =  m^2 q^{\star 2} \bigg(\frac{g^2}{m_{R}^2 - s}+h(s) \bigg) \,,
\label{eq:Kmatpar} 
\end{gather}
where $g$ is a dimensionless coupling constant, $m_R$ is a parameter with units of energy, and $h(s)$ is a polynomial in $s$ with dimensions $1/m^2$. The dimensions of these parameters are chosen such that 
$\K $ has dimensions of $m^2$. For the transition form factor, $\mathcal{A}$, and the smooth function $\mathbf{S}$ we choose:
\begin{equation}
\mathcal{A}(s, Q^2) = 
\frac{1}{1+{Q^2}/m_R^2} \,, \qquad
\mathbf{S}(s,Q^2,Q_{if}^2) = 0 \,.
\label{eq:Hparam}
\end{equation}

\subsubsection*{Naive analysis}

\begin{figure}[t!]
\begin{center}
\includegraphics[width=1\textwidth]{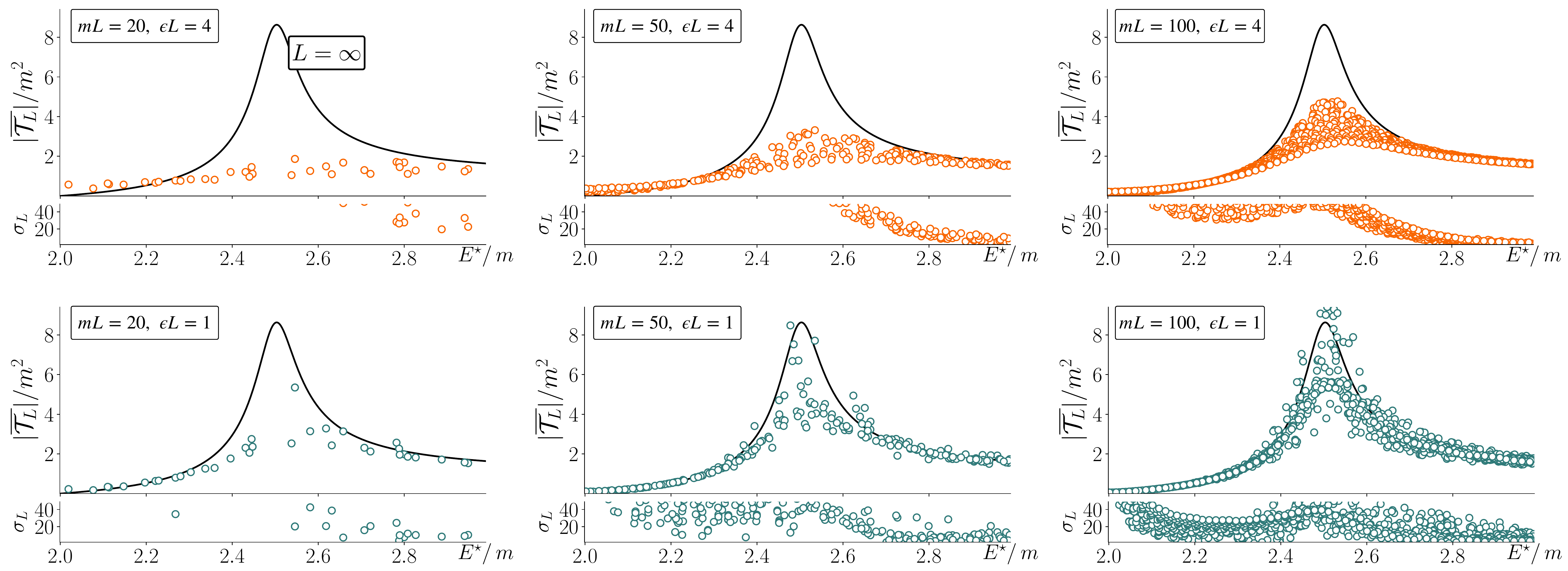}
\caption{Infinite-volume amplitude, $\T$ (black curve) vs. finite-volume estimator $\overline {\T_L}$ (defined in Ref.~\cite{Briceno:2020rar}) (colored points), for a single channel open. The photon virtualities are $Q^2=Q^2_{if}=2 m^2$, and the binning resolution is $\Delta_{Q^2}=0.01 m^2$ (see Eq.~\eqref{eq:BinSet1}). The smaller plots below each panel represent the percent deviation, quantified by with $\sigma_L$ defined in Eq.~\eqref{eq:sigmaL}.  \label{fig:iTL_disc}} 
\end{center}
\end{figure}

Using the parametrization above for the K matrix, we consider a set of resonant amplitudes given by $m_R = 2.5m$, $g=2.5$ and $h(s)=0$. We then evaluate the finite volume dependence of $\mathcal{T}_L$ numerically in Figure~\ref{fig:iTL_disc} for three different values of $L$, $mL = 20,\, 50,\, 100$ and two values of $\epsilon L = 1,\,4$. The black lines represent the infinite volume Compton amplitude, $\mathcal{T}$, while the colored dots represent an estimator for the finite volume scattering amplitude analog denoted by $\overline {\T_L}$ (for details see Section~IV.B in Ref.~\cite{Briceno:2020rar}). This estimator, $\overline {\mathcal T_L} $, is computed within a suitable kinematic bin defined by,
\begin{equation}
\big |\overline{Q^2}-{Q}^2\big| < \Delta_{Q^2} \qquad \text{and} \qquad \big|{Q}_{if}^2-{Q}^2\big| < \Delta_{Q^2} \,,
\label{eq:BinSet1}
\end{equation}
where we fixed the target virtuality $\overline{Q^2} = 2m^2$. We also fix the virtuality resolution to $\Delta_{Q^2}=0.01 m^2$. The deviation from $\mathcal{T}_L(E+i\epsilon)$ to $\mathcal{T}(E^\star)$ can be quantified using, 
\begin{align}
\sigma_L(E^\star, \boldsymbol P, \epsilon)=100\times\left|
\frac{\T_L(E + i \epsilon, \boldsymbol P)-\T(E^\star)}{\T(E^\star)}
\right| \,,
\label{eq:sigmaL}
\end{align}
plotted in the panels below $\overline {\mathcal T_L}/m^2$. From Figure~\ref{fig:iTL_disc} we note that in general $\overline {\mathcal T_L}$ shows substantial deviations from the infinite volume amplitude, in particular around the peak of the amplitude, which can be attributed to a nearby resonance. Only for volumes as large as $mL = 100$ and $\epsilon$ as small as $\epsilon L = 1$ these deviations are reduced but still far from the percent level. 

\subsubsection*{Boost averaging}

As discussed with detail in Ref.~\cite{Briceno:2020rar}, the scenario shown in Figure~\ref{fig:iTL_disc} can be improved by exploiting the fact that $\mathcal{T}$ depends only on Lorentz scalars, while  $\mathcal{T}_L$ depends on the total momentum of the system. Therefore, binning and averaging over similar kinematic points makes the finite volume estimator $\overline {\T_L}$ to converge faster to the physical amplitude. Also, we consider an average over several volumes, since this largely cancel the fluctuations associated with a single value of $L$. To perform this average we sample ${\T_L}$ in bins centered at a fixed value $\overline E^\star$, each bin with a width $\Delta_{E^\star}$. We then average all the values of $\T_L$ lying in the 3D-bin defined by:
\begin{equation}
\big |\overline{Q^2}-{Q}^2\big| < \Delta_{Q^2}\, , \qquad \qquad \big |{Q}_{if}^2-{Q}^2\big| < \Delta_{Q^2} \qquad \text{and} \qquad |\overline {E^\star} - E^\star| \leq \Delta_{E^\star} \,. 
\label{eq:BinSet2}
\end{equation}

\begin{figure}[t!]
\begin{center}
\includegraphics[width=1\textwidth]{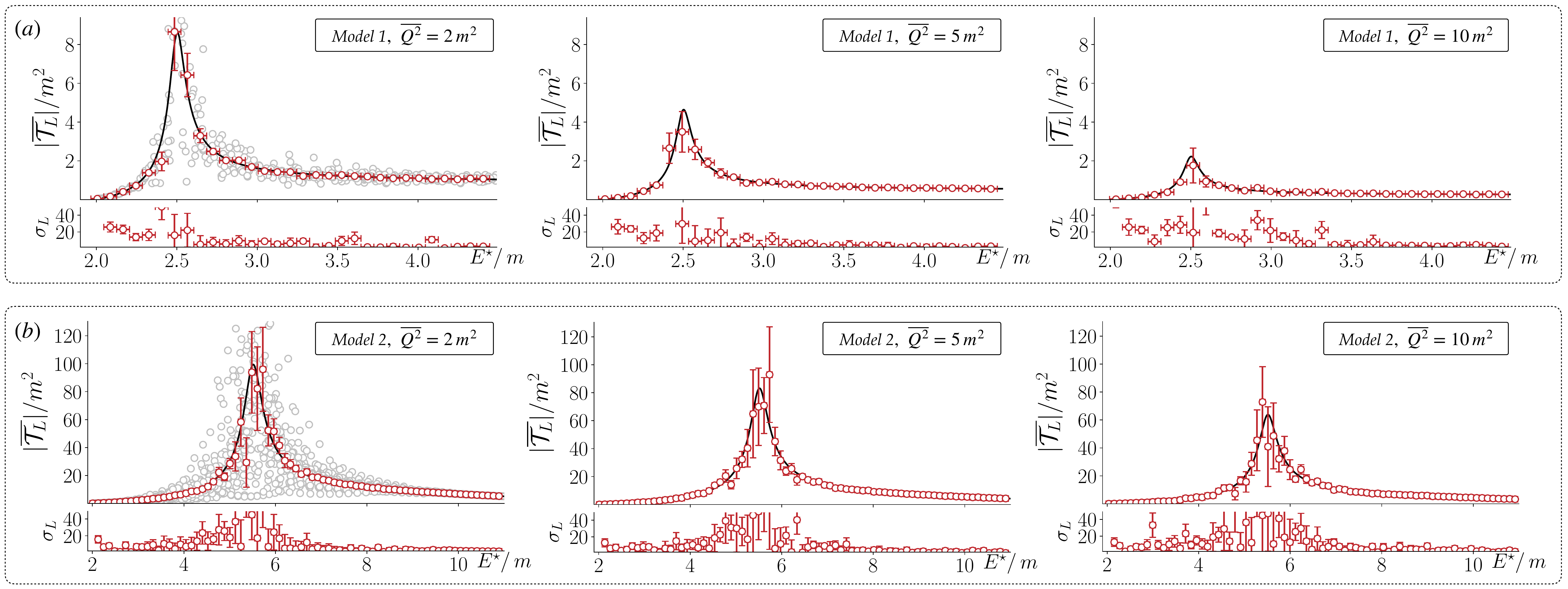}
\caption{ 
Infinite-volume amplitude, $\T$ (black curve) vs. the finite-volume estimator (defined in Ref.~\cite{Briceno:2020rar}) (red points) for a single channel open $(a)$ Data is generated using the \emph{Model 1} set of parameters (used in Figure.~\ref{fig:iTL_disc}): $m_R=2.5m$, $g=2.5$, $h(s) = 0$. $(b)$ Data is generated using  \emph{Model 2} set of parameters: $m_R=5.5m$, $g=6$, $h(s) = 0.2/m^2$. The light grey points in the two panels on the left correspond to the values of $\mathcal T_L$ obtained from points with similar kinematics (see Eq.~\eqref{eq:BinSet2}). These light-gray points are then used to compute $\overline{\mathcal T_L}$. Although the formalism used strictly holds only below the three-particle threshold, we take the liberty to extend to energies well above this threshold. 
\label{fig:iTL_binning}}
\end{center}
\end{figure}

In the upper panels of Figure~\ref{fig:iTL_binning}, we consider the {\it Model 1} defined by $m_R = 2.5m$, $g=2.5$ and $h(s)=0$ and compute the average $\overline \T_L$ over three different volumes, $mL = 20\,, 25\,, 30$, for three target virtualities, $\overline{Q^2} = 2m^2, 5m^2, 10m^2$.  The size of the energy bins is $\Delta_{E^\star} = 0.08m$, while the virtuality bins have a size of $\Delta_{Q^2}=0.05\,m^2$ and our $\epsilon$ choice is $\epsilon(L)=1/[L (mL)^{1/2}]$. In the bottom panels of Figure~\ref{fig:iTL_binning} we consider the {\it Model 2} defined by $m_R=5.5m$, $g=6$, $h(s) = 0.2/m^2$. In both cases, we note that the proposed averaging provides the best reconstruction to the infinite volume Compton Amplitude.

\subsection{Multiple open channels}
\label{sec:Multiple_channels_numerical}

\begin{figure}[ht]
\centering
\subfigure{\includegraphics[width=0.32\linewidth]{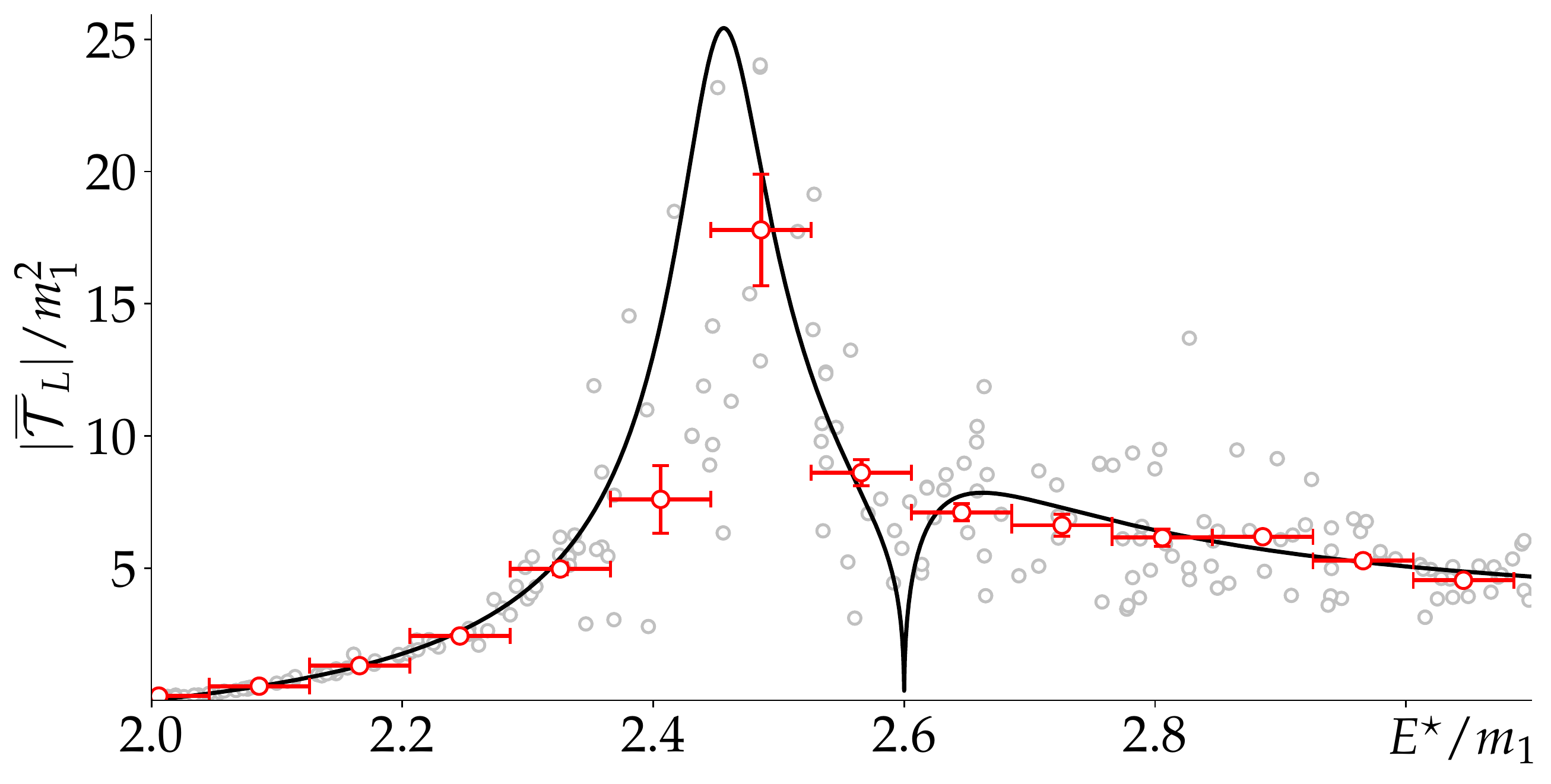}}
\subfigure{\includegraphics[width=0.32\linewidth]{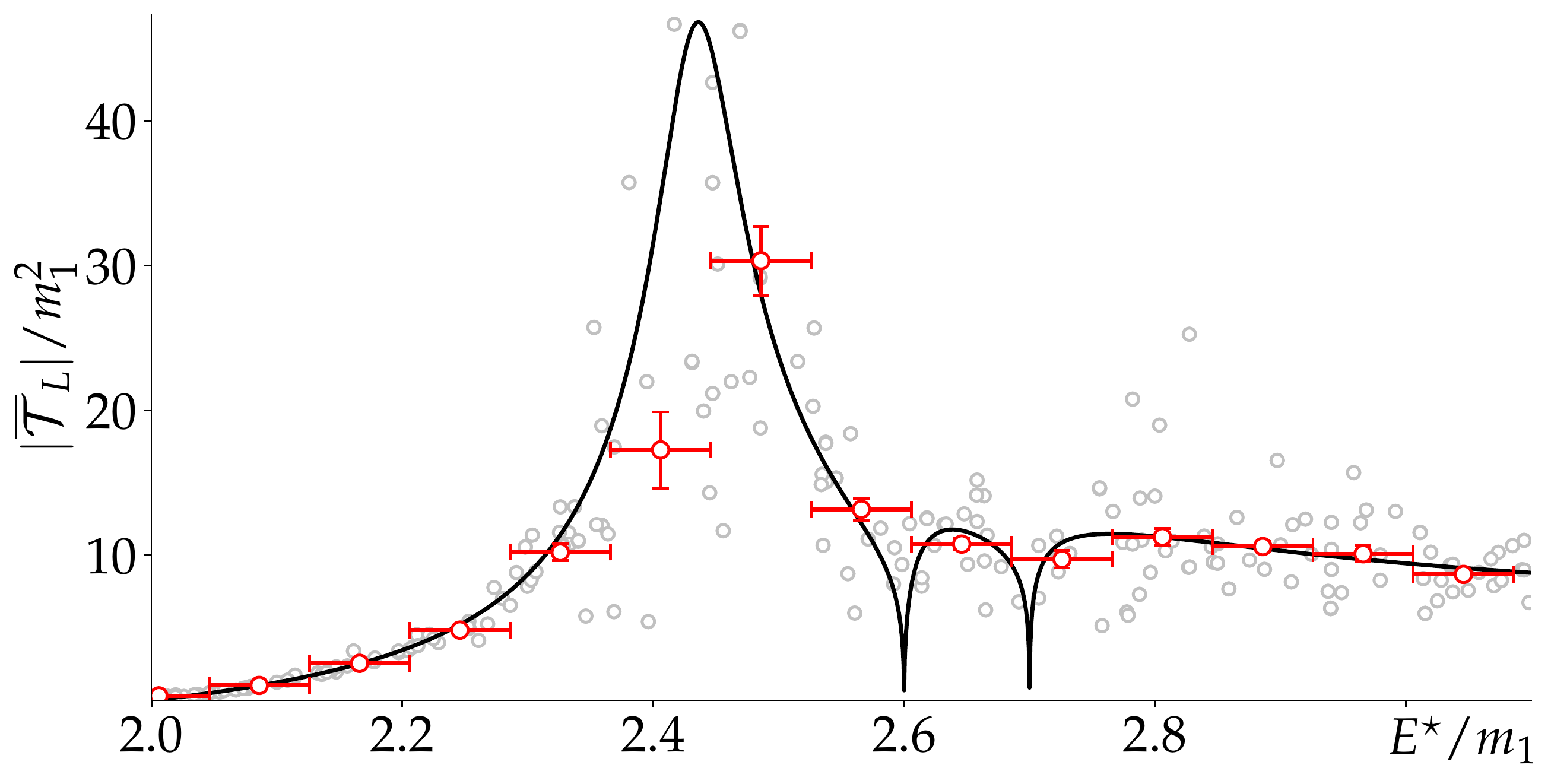}}
\subfigure{\includegraphics[width=0.32\linewidth]{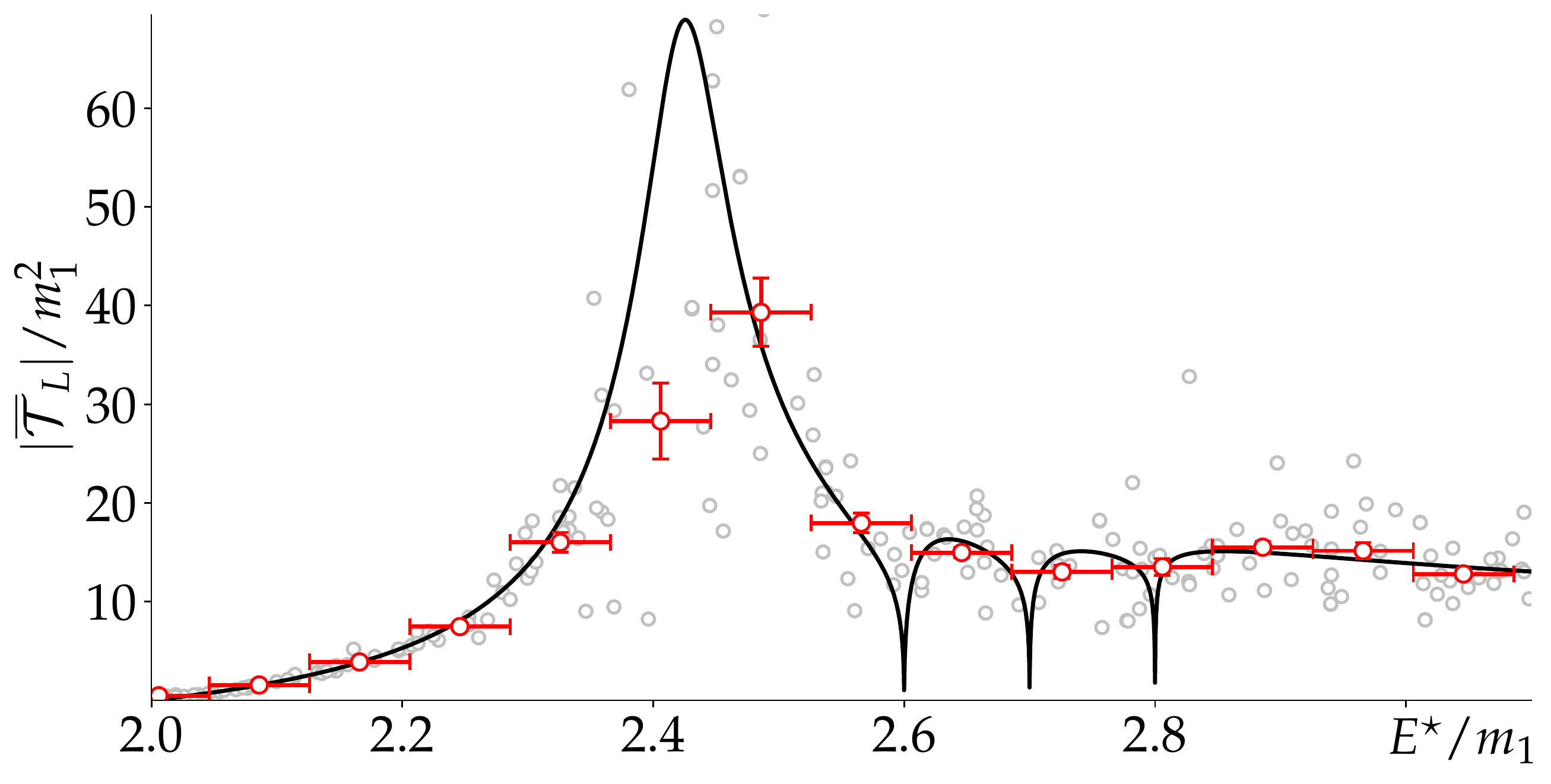}}
\subfigure{\includegraphics[width=0.32\linewidth]{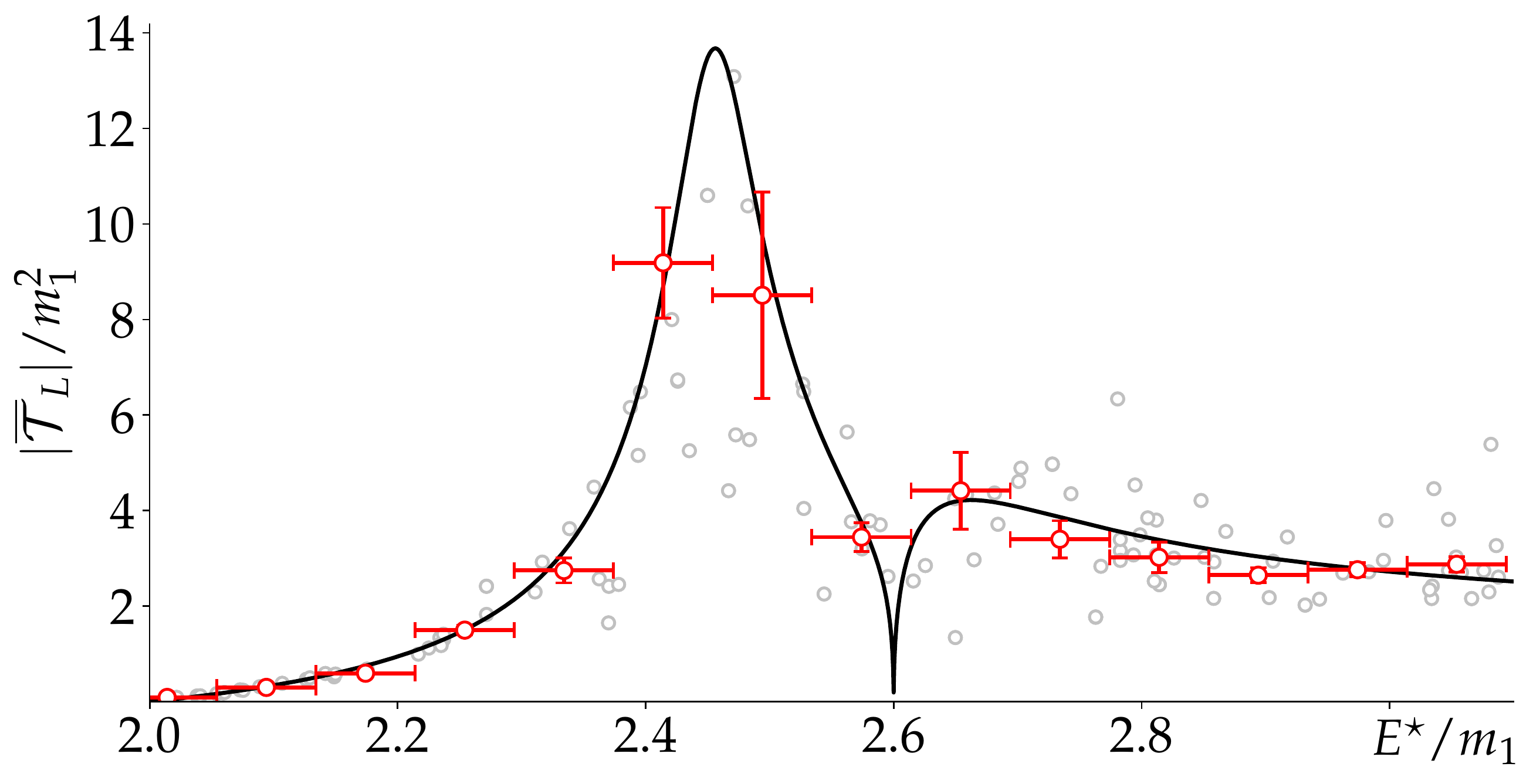}}
\subfigure{\includegraphics[width=0.32\linewidth]{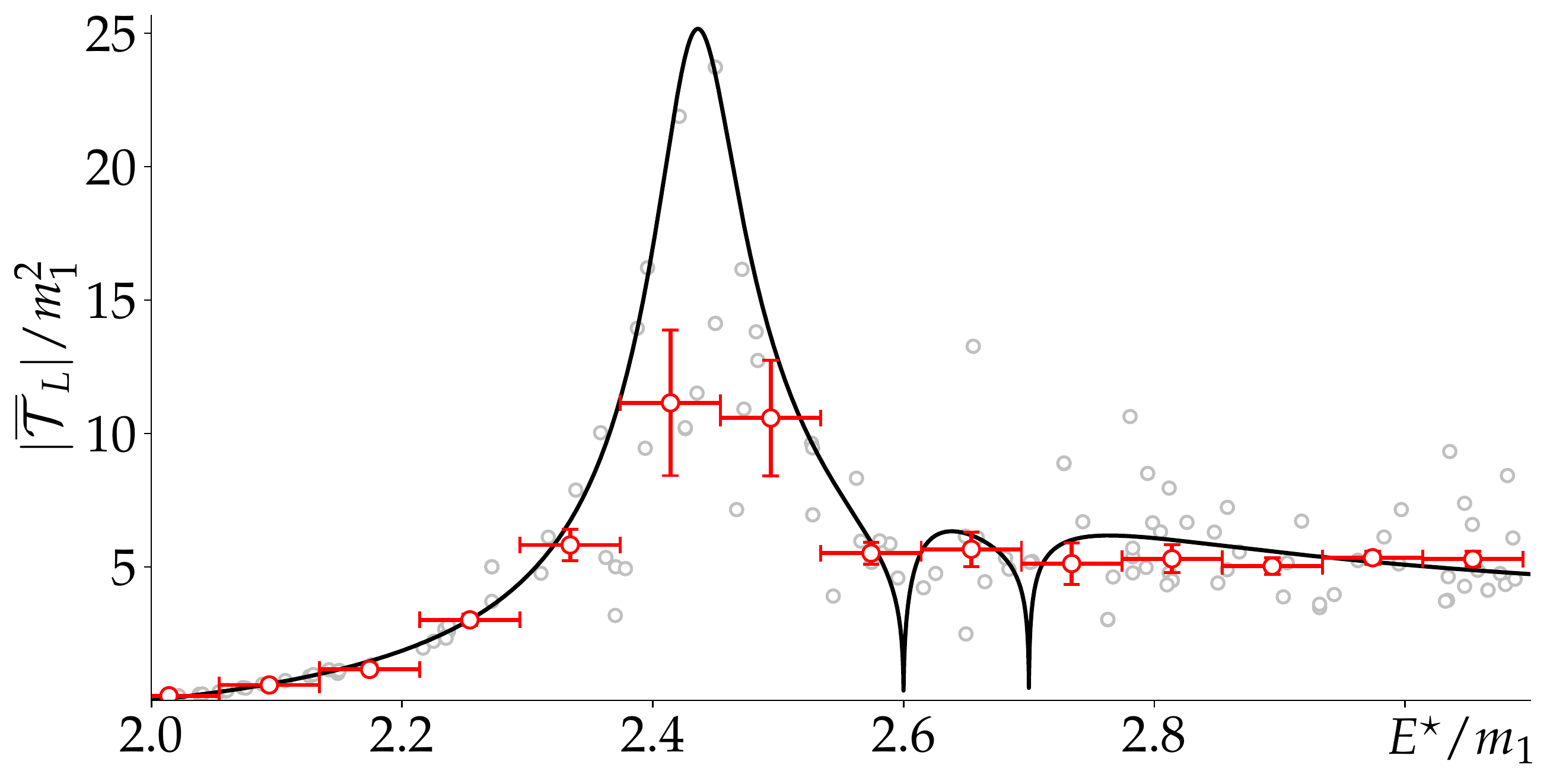}}
\subfigure{\includegraphics[width=0.32\linewidth]{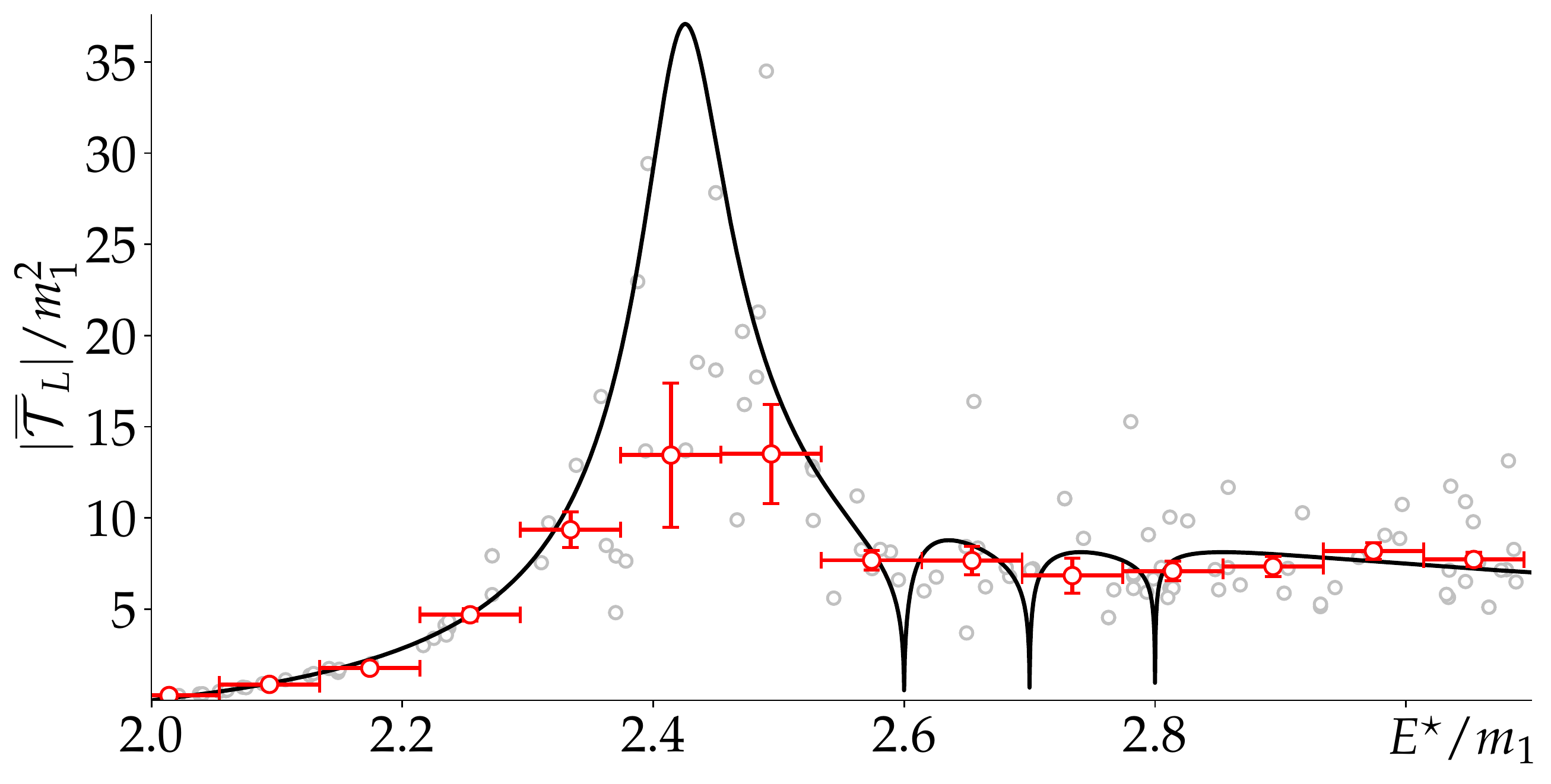}}
\caption{The red points represent the binned Compton amplitude obtained from the points with similar kinematics, shown in gray, for volumes $L=20,$ $25,$ and $30$. The black solid line is the infinite volume Compton amplitude. From left to right there are 2, 3, and 4 open channels with masses $m_2=1.3m_1$, $m_3=1.35m_1$, and $m_4=1.4m_1$ and coupling constants $g_1=2.5$, $g_2=1.5$, $g_3=1.35$, and $g_4=0.985$. Here we consider $\Delta_{Q^2}=0.05m_1^2$ and $\Delta_{E^\star}=0.08m_1$ for the binning conditions, as well as the smooth functions $\mathbf{S}(s,Q^2, Q_{if}^2)=0$, and $h_{ab}(s)=0$. The incoming and outgoing virtualities are $Q^2=Q_{if}^2=2m_1^2$, for the top, and $Q^2=Q_{if}^2=5m_1^2$ for the bottom.}  
\label{Fig:CC_Bin_Compton}
\end{figure}

In Ref.~\cite{Briceno:2020rar} we outlined a procedure for accessing the Compton amplitude given arbitrary values of $s$. However, evidence that this procedure works was only shown explicitly for kinematics where a single channel composed of two particles may go on-shell. In this section we provide preliminary empirical evidence that these observations persist even for kinematics where multiple two-body channels may go on-shell. In this case, our parametrization for the K-matrix is given by a simple generalization of Eq.~\eqref{eq:Kmatpar}
\begin{align}
\mathcal{K}_{ab}(s) 
    &= m_a m_bq_1^{\star 2}\left( 
         \frac{g_a g_b}{m_R^2-s} 
         + h_{ab}( s )\right),
\end{align}
where $g_a$ is the coupling constant to the $a$-th channel and $h_{ab}(s)$ is a matrix whose elements are polynomials in $s$. While the transition form factors we fix to
\begin{align}
\mathcal{A}_a(s,Q^2) 
    &= \frac{1}{1+Q^2/m_R^2}. 
\end{align}
For equal incoming and outgoing virtualities, $Q^2 = Q^2_{if}$, binning conditions $\Delta_{Q^2}=0.05m_1^2$ and $\Delta_{E^\star}=0.08m_1$, smooth function $\mathbf{S}_{ab} (s,Q^2,Q_{if}^2)=0$, and matrix $h_{ab}(s)=0$, we find the results shown in Fig.~\ref{Fig:CC_Bin_Compton} considering two, three, and four open channels with corresponding masses $m_2=1.3m_1$, $m_3=1.35m_1$, and $m_4=1.4m_1$ and coupling constants $g_1=2.5$, $g_2=1.5$, $g_3=1.35$, and $g_4=0.985$. These results support the hypothesis that the method outlined in Ref.~\cite{Briceno:2020rar} holds for an arbitrary number of open channels.

\section{Final Remarks}

In this work we have explored the prospects of accessing Compton-like amplitudes in real-time calculations of a 1+1-dimensional theory with periodicity $L$ in the single spatial direction. A finite-volume, non-zero $\epsilon$ estimator for the Compton amplitude can be defined, which coincides with the physical amplitude in the ordered double limit: first $L \to \infty$ followed by $\epsilon \to 0$. Having defined this quantity, the practical issue arises of whether values of $\epsilon$ and $L$ can be identified to give a predicted value that is not dominated by systematic uncertainties.

To explore this question we have taken the formalism of Ref.~\cite{Briceno:2019opb} for extracting finite-volume long-range matrix elements as a diagnostic tool. It is worth stressing that the formalism is used here in a manner completely distinct from the main focus of that work. Instead of using finite-volume information from lattice QCD calculations to predict infinite-volume amplitudes, here we take an ansatz for the infinite-volume amplitudes to predict the finite-volume, non-zero-$\epsilon$ estimator. This allows us to quantify finite-volume effects that might be seen by future real-time simulations, assuming that the latter do not make use of the formalism of Ref.~\cite{Briceno:2019opb}.

For the systems we consider, in particular those with a resonant peak of width comparable with typical low-lying QCD resonances, the value of $\epsilon$ must be taken sufficiently small to not distort the amplitude. But taking values in the regime where the $\epsilon \to 0$ extrapolation is feasible, we find that
finite-volume effects become significant, to the extent that one requires volumes of order $mL = \mathcal{O}(10^2)- \mathcal{O}(10^3)$ to reduce these systematics to the $5-10\, \%$ level. We present a practical solution to overcome this issue which relies on  exploiting symmetries of the infinite-volume amplitudes, binning over similar kinematics and averaging over each bin. The proposed average converges faster to the infinite-volume amplitude and requires volumes of order $mL = 20- 30$. Here we provide first evidence that this procedure also works for kinematics in which two or more two-body channels are kinematically open.

\section{Acknowledgments}
RAB and JVG are partly supported by the USDOE grant under which Jefferson Science Associates, LLC,\, manages and operates Jefferson Lab, \, No. DE-AC05-06OR23177. Additionally, RAB acknowledges support from the USDOE Early\, Career award, contract de-sc0019229. MCB and JVG are also supported by the Jefferson Lab LDRD project LD2117. MTH is supported by UK Research and Innovation Future Leader Fellowship MR/T019956/1, and also in part by UK STFC grant ST/P000630/1.

\bibliographystyle{JHEP}
\bibliography{References}{}

\end{document}